\documentstyle[prd,aps,floats,epsfig]{revtex}

\newcommand{\beq}{\begin{equation}}
\newcommand{\eeq}{\end{equation}}
\newcommand{\bea}{\begin{eqnarray}}
\newcommand{\eea}{\end{eqnarray}}

\def\laq{~\raise 0.4ex\hbox{$<$}\kern -0.8em\lower 0.62
ex\hbox{$\sim$}~}
\def\gaq{~\raise 0.4ex\hbox{$>$}\kern -0.7em\lower 0.62
ex\hbox{$\sim$}~}

\def \pa {\partial}
\def \ra {\rightarrow}
\def \la {\lambda}

\def \b {\beta}
\def \a {\alpha}

\def \da {\delta}

\def \r {\rho}

\def \Om {\Omega}
\def \noi {\noindent}

\begin{document}

\def\baselinestretch{1.4} 

{\large
\begin{flushright}
BA-TH/00-389\\
July 2000\\
astro-ph/0009476
\end{flushright}
}

\vspace*{0.3truein}

\vskip 1.3 cm
{\Large\bf\centering\ignorespaces
New mechanism for the generation of primordial seeds\\
for the cosmic magnetic fields
\vskip2.5pt}
{\dimen0=-\prevdepth \advance\dimen0 by23pt
\nointerlineskip \rm\centering
\vrule height\dimen0 width0pt\relax\ignorespaces

\vskip 1 cm
{\large  M. Gasperini}
\par}
{\large\it\centering\ignorespaces
Dipartimento di Fisica, Universit\`a di Bari, \\
Via G. Amendola 173, 70126 Bari, Italy\\
and \\
Istituto Nazionale di Fisica Nucleare, Sezione di Bari,
Bari, Italy \\
\par}

\par
\bgroup
\leftskip=0.10753\textwidth \rightskip\leftskip
\dimen0=-\prevdepth \advance\dimen0 by17.5pt \nointerlineskip
\small\vrule width 0pt height\dimen0 \relax
 
\vskip 1.5 cm
\centerline{\Large Abstract}
\vskip 0.5 cm
\noi
{\large 
We discuss the inflationary production of magnetic seeds for the
galactic dynamo through the photon-graviphoton mixing, typical of
extended models of local supersymmetry. An analysis of the allowed
region in parameter space shows that such a mechanism is compatible
with existing phenomenological bounds on the vector mass and mixing
parameter, but requires a large enough mixing that seems difficult
to  implement naturally in the context of a  minimal supersymmetric
scenario. 
}

\par\egroup

{\large 
\vspace{0.8cm}
\begin{center}
------------------------------  

\vspace{0.5cm}
To appear  in {\bf  Phys. Rev. D}
\end{center}
}

\vfill


\def\baselinestretch{1}
\newpage
\setcounter{page}{1}

\par
\begingroup
\twocolumn[%

\begin{flushright}
BA-TH/00-389\\
astro-ph/0009476\\
To appear in {\bf Phys. Rev. D}
\end{flushright}
\bigskip

{\large\bf\centering\ignorespaces
New mechanism for the generation of primordial seeds\\
for the cosmic magnetic fields
\vskip2.5pt}
{\dimen0=-\prevdepth \advance\dimen0 by23pt
\nointerlineskip \rm\centering
\vrule height\dimen0 width0pt\relax\ignorespaces
 M. Gasperini 
\par}
{\small\it\centering\ignorespaces
Dipartimento di Fisica, Universit\`a di Bari, 
Via G. Amendola 173, 70126 Bari, Italy\\
and Istituto Nazionale di Fisica Nucleare, Sezione di Bari, Bari,
Italy 
\par}

\par
\bgroup
\leftskip=0.10753\textwidth \rightskip\leftskip
\dimen0=-\prevdepth \advance\dimen0 by17.5pt 
\nointerlineskip
\small\vrule width 0pt height\dimen0 \relax

We discuss the inflationary production of magnetic seeds for the
galactic dynamo through the photon-graviphoton mixing, typical of
extended models of local supersymmetry. An analysis of the allowed
region in parameter space shows that such a mechanism is compatible
with existing phenomenological bounds on the vector mass and mixing
parameter, but requires a large enough mixing that seems difficult
to  implement naturally in the context of a  minimal supersymmetric
scenario. 

\par\egroup
\vskip2pc]
\thispagestyle{plain}
\endgroup

The parametric amplification of the vacuum fluctuations \cite{1},
induced by inflation, is one of the most appealing mechanisms \cite{2}
for the spontaneous generation of the large-scale magnetic fields
required to seed the galactic dynamo, or the galactic magnetic field
itself \cite{3}. Unfortunately, the minimal coupling of photons to a
four-dimensional geometry is conformally invariant: in that case, the
electromagnetic fluctuations are unaffected by the time evolution of a
conformally flat metric (typical of the inflationary scenario) and then,
in particular, cannot be amplified. 

Up to now, known attempts to generate large enough magnetic seeds
from the vacuum try to break conformal invariance either at the
classical level, through some {\sl ad-hoc} nonminimal coupling of
photons to the background curvature \cite{2}, or at the quantum leel,
though the so-called ``trace-anomaly" \cite{4} (or even through a
spontaneous breaking of Lorentz invariance \cite{4a}).  
Alternatively, it is possible to exploit the nonminimal coupling of
photons to a scalar field, the inflaton \cite{5} or the dilaton \cite{6},
which cannot be eliminated by a conformal rescaling of the metric, and
which plays the role of the external ``pump field" amplifying the
electromagnetic fluctuations. The coupling to a dynamical dilaton, in
particular, is naturally provided by superstring theory, and may act
efficiently in the context of the pre-big bang scenario \cite{6}. 

The physical origin of the conformal invariance, which prevents the
inflationary amplification of minimally coupled electromagnetic
fluctuations, is the unbroken gauge symmetry of the Maxwell action,
leading to massless photons. Even if massless, however, the photon
field $A_\mu$ could be non-trivially mixed with the massive vector
component $V_\mu$ of the gravitational supermultiplet -- the
so-called ``graviphoton" -- according to the effective action: 
\bea
&&
S = \int {d^4x\over 8\pi}\sqrt{-g}~(-{1\over 2}F_{\mu\nu}F^{\mu\nu} 
-{1\over 2}G_{\mu\nu}G^{\mu\nu} \nonumber\\
&&
~~~~~~~+\a F_{\mu\nu}G^{\mu\nu}+m^2
V_\mu V^\mu), 
\label{1}
\eea
where $F_{\mu\nu}= 2\pa_{[\mu }A_{\nu]}$,  $G_{\mu\nu}= 2\pa_{[\mu
}V_{\nu]}$. Such a gauge-invariant mixing, parametrized by the
dimensionless coefficient $\a$ ($|\a|<1$ to avoid unphysical states in
the mass spectrum), is indeed required by local supersymmetry in $N=2$
extended models for the graviphoton \cite{7}, and may also arise
naturally in the context of higher-dimensional gauge interactions
\cite{8}. The above action also describes, formally, the effective mixing
of photons and paraphotons \cite{9}, and is phenomenologically
equivalent to a particular case of the two-photon electrodynamics
earlier studied by Okun \cite{10}. 

The mass of the $V_\mu$ field breaks the conformal invariance of the
action and makes possible, in principle, the inflationary amplification of
vector fluctuations. We note, for future reference, that in the
context of supersymmetric models the vector mass $m$ is typically of
the order of the gravitino mass \cite{7}, and thus expected to be fixed
by SUSY breaking mechanisms at the natural scale \cite{7a} $m \sim 1$
TeV (see however \cite{7b} for the compatibility of smaller masses with
realistic models of supersymmetry breaking). In the context of
higher-dimensional models, on the contrary, the vector mass can be
much lighter \cite{8a}, so as to match the (phenomenologically
interesting) macroscopic range $1$ m -- $1$ Km, provided new
supersmall mass scales are introduced \cite{8}. Also, in supersymmetric
models the mixing with the photon is expected to be strongly
suppressed by the fundamental (quark or lepton) mass scale $M$
expressed in Planck units \cite{7}, so that one obtains, typically, a very
small mixing parameter, $\alpha \sim M/M_p \sim 10^{-20}$. The mixing
can be much larger, however, in a higher -dimensional context \cite{8},
if the vector mass is fixed (for instance) around the weak interaction
scale.

Quite irrespective of the physical (supersymmetric and/or
higher-dimensional) origin of the mixing, and of the particular
mechanism of mass generation for the graviphoton, the aim of this
paper is to discuss whether, for phenomenologically allowed values of
$\a$ and $m$, the fluctuations of the massive vector can be efficiently
amplified (because of the mass) by inflation, and efficiently converted
(through the mixing) into electromagnetic fluctuations, strong enough
to seed the cosmic magnetic fields observed on (inter)galactic scale.  It
should be recalled, to this purpose, that if we identify $V_\mu$ with
the so-called ``fifth-force" vector boson \cite{11}, coupled to baryon
number, then the strength of the mixing is significantly constrained by
geomagnetic data and by Cavendish's tests of Coulomb's law  in
the geophysical mass range \cite{12}, and by pion and kaon decays in
the nuclear mass range \cite{13}. Independently from the
possible sources of $V_\mu$,  the mixing is also
constrained by the induced photon-graviphoton oscillations in a
dielectric \cite{14}, and by changes of electromagnetic polarization due
to a possible anomalous magnetic moment of the massive vector
\cite{15}. 

In order to discuss the cosmological amplification of the vacuum
fluctuations, in a conformally flat metric $g_{\mu\nu}$, it is convenient
to use the conformal time $\eta$ by setting $g_{\mu\nu}= a^2(\eta)
\eta_{\mu\nu}$, and to impose the gauge conditions $\pa_\mu
A^\mu=0=\pa_\mu V^\mu$. The action (\ref{1}) then becomes, after
integration by part, 
\bea
&&\qquad
S= \int {d^3x d\eta\over 8\pi}\Bigg[A_\mu \Box A^\mu+   V_\mu \Box
V^\mu 
\nonumber\\
&&\qquad
- 2\a A_\mu \Box V^\mu + m^2 a^2(\eta) V_\mu V^\mu\Bigg],
\label{2}
\eea
where the vector indices are now raised and lowered with the
Minkowski metric $\eta_{\mu\nu}$, and $\Box \equiv \pa^2_\eta -
\nabla^2$ is the flat-space Dalembertian in the conformal time gauge. 
In this gauge, the coupling to the background geometry  only survives
in the mass term, through the scale factor $a(\eta)$. 

The above action describes the vector mixing of a non-orthogonal
combination of the two mass eigenstates, $\psi_\mu, \phi_\mu$,
coupled to a time-dependent external field, $a(\eta)$. By setting
\beq
\pmatrix{A^\mu \cr V^\mu \cr}= \pmatrix{1 & \a \cr 0 & 1 \cr} 
\pmatrix{\psi^\mu \cr \phi^\mu \cr}
\label{3}
\eeq
the action takes indeed the diagonal form 
\beq
S= \int {d^3x d\eta\over 8\pi}\Bigg[\psi_\mu \Box \psi^\mu
+(1-\a^2) \phi_\mu \left(\Box +{m^2 a^2\over 1- \a^2}\right) \phi^\mu
\Bigg],
\label{4}
\eeq
and leads to the evolution equations
\bea
&&
\left(\psi^\mu_k\right)'' + k^2 \psi^\mu_k=0, \\
&&
\left(\phi^\mu_k\right)'' + \left(k^2 +{m^2 a^2 \over 1-\a^2} \right)
\phi^\mu_k=0. 
\label{6}
\eea
Here the prime denotes differentiation with respect to  $\eta$, and we
have expanded the fields into Fourier modes, $\nabla^2 \psi_k = -k^2
\psi_k$. Given an initial vacuum fluctuation spectrum, $\da
\psi_k^\mu$, $\da \phi_k^\mu$, the massless fluctuations evolve freely,
unaffected by the background geometry, and only the massive
components, with effective mass $ \overline m = m/\sqrt{1-\a^2}$, are
amplified. After the amplification we can thus set $\da \psi^\mu_k=0$,
and the spectrum of the electromagnetic fluctuations, induced by the
mixing of eq. (\ref{3}), turns out to be proportional to the massive
vector spectrum, $\da A^\mu_k= \a \da \phi^\mu_k$. The spectral
energy densities, $ \rho(k)= d \rho /d \ln k$, are quadratic in the fields, 
and are thus related by: 
\beq
\r_A(k) = \a^2 \r_\phi(k)= \a^2 \r_V(k). 
\label{7}
\eeq

For a first, indicative estimate of the possibility of seed production we
shall now compute the massive vector spectrum $\r_\phi(k)$, following
the standard methods of cosmological perturbation theory \cite{16},
and assuming a ``minimal" model of cosmological evolution in which
the standard radiation era, starting at $\eta=\eta_1$, is preceeded in
time by a phase of exponentially expanding, de Sitter-like inflation
(there is no need of considering also the transition to the
matter-dominated epoch since, for our purpose, it will be enough to
evaluate the spectrum at the intergalactic scale $k/a\sim (1 {\rm
Mpc})^{-1}\sim 10^{-14}$Hz, which re-enters the horizon before the
matter-radiation equilibrium).  Thus we set $ a(\eta) = -
(H_1\eta)^{-1}$  for $\eta \ll \eta_1$, and $ a \sim \eta$ for $\eta \gg 
\eta_1$. Our final electromagnetic spectrum will be characterized by
three free parameters: the graviphoton mass $m$, the mixing
coefficient $\a$, and the inflation scale $H_1$. 

In the initial inflationary phase eq. (\ref{6}), for each physical
polarization modes $\phi_k$, in our case  reduces to 
\beq
\phi_k'' +\left(k^2 +{\overline m^2\over H_1^2 \eta^2}\right)\phi_k =0, 
~~~~~~~ \overline m = {m\over \sqrt{1-\a^2}},
\label{8a}
\eeq
and the solution, normalized at $\eta \ra -\infty$ to a vacuum
fluctuation spectrum \cite{16}, can be written in terms of the 
second kind Hankel functions \cite{17} as
\beq
\phi_k(\eta)= |\eta|^{1/2}H_\nu^{(2)} (|k\eta|), 
~~ \nu = {1\over 2}\left(1-4{\overline m^2\over
H_1^2}\right)^{1\over 2}, ~\eta <\eta_1.
\label{8}
\eeq
We are interested in modes which are non-relativistic ($k<\overline m
a$) already at the transition epoch $\eta_1$, since for higher modes
the mass term can be neglected in eq. (\ref{6}), and there is no
significant amplification throughout the whole inflationary epoch. Such
modes are defined by the condition $k<k_M$, where $k_M=\overline m
a_1 =(\overline m /H_1)k_1$, and $k_1 = |\eta|^{-1}$ is the limiting
mode crossing the horizon just at $\eta=\eta_1$. We shall assume that
the graviphoton mass is much smaller than the inflation scale
(we have in mind, typically,  $\overline m \simeq m \sim 1$ TeV), so that
$k_M/k_1 = (\overline m/H_1) \ll1$. 

In the subsequent radiation era the mass term $ma(\eta)$ grows linearly
in conformal time, and then dominates always better the evolution
equation for all non-relativistic modes $k<k_M$. For such modes we can
thus approximate eq. (\ref{6}), in the radiation era, by
\beq
\phi_k''+\b^2 \eta^2 \phi_k=0, ~~~~ \b=\overline m H_1 a_1^2, ~~~~
k<k_M,
\label{9}
\eeq
and the general solution can be written in terms of the Hankel
functions as 
\bea
&&
\phi_k(\eta)= \eta^{1/2}\left[c_+(k)H_{1/4}^{(2)}
\left(\b\eta^2\over 2\right)+c_-(k)H_{1/4}^{(1)} \left(\b\eta^2\over
2\right)\right], \nonumber \\
&&
~~~~~~~~~~~~~~k<k_M, ~~~~~~~~\eta >\eta_1.
\label{10}
\eea
Here $c_\pm(k)$ are the so-called Bogoliubov coefficients \cite{16}, 
determining the spectral number distribution of the pair of vector
particles produced from the vacuum. They can be computed by matching
the two solutions (\ref{8}) and (\ref{10}) and their first derivative at
$\eta=\eta_1$. Such a matching can be easily performed by exploiting
the small argument limit of the Hankel functions (since
$|k\eta_1|<k_M/k_1=\overline m/H_1=\b \eta_1^2 \ll1$), and leads to 
\beq
|c_-(k)|\simeq |c_+(k)| \simeq \left(k\over
k_1\right)^{-\nu}\left(H_1\over \overline m\right)^{1/4}, ~~~k <k_M
\label{11}
\eeq
(here and throughout we will neglect numerical factors of order unit,
as we are primarly interested in an order of magnitude estimate  of
seed production). We may note that $|c_\pm (k)|\gg1$, signalling the
effective entry of non-relativistic modes into the parametric
amplification regime. 

The spectral energy distribution of the non-relativistic vector
fluctuations, amplified by inflation, can now be conveniently expressed
in terms of the proper wave-number $p=k/a$, and referred to the total
critical energy density $\r_c=3M_P^2H^2/8\pi$ ($M_P$ is the Planck
mass) as
\bea
&&
\Om_\phi(p,t) \equiv {p\over \r_c}{d\r_\phi\over dp}\simeq 
{\overline m p^3 |c_-(p)|^2\over M_P^2H^2}\simeq \nonumber\\
&&
\left(H_1\over M_P\right)^2 \left(\overline m \over
H_1\right)^{1\over 2} \left(H_1\over H\right)^2
 \left(a_1\over a\right)^3  \left(p\over p_1\right)^2,
~p<p_M,
 \label{12}
\eea
where we have set $\nu=1/2$ from eq. (\ref{8}), 
working in the assumption $\overline m/H_1 \ll1$. Here $p_M=
(\overline m/H_1 )p_1$, and $p_1=k_1/a_1=H_1a_1/a$ is the proper
momentum scale crossing the horizon at the end of inflation (today
$p_1(t_0)\simeq (H_1/M_P)^{1/2} 10^{11}$ Hz). During the
matter-dominated era, the above spectrum keeps frozen at the
value reached at the time of equilibrium,
$\Om_\phi(t_0)=\Om_\phi(t_{\rm eq})$. On the other hand, using 
the kinematics of the radiation era, $(H_1/H_{\rm eq})^2(a_1/a_{\rm
eq})^3$=$(H_1/H_{\rm eq})^{1/2}$. The final energy spectrum of the
electromagnetic fluctuations, obtained from the vacuum through the
photon-graviphoton mixing, can be finally estimated from eqs. (\ref{7})
and (\ref{12}) as follows:
\beq
\Om_A(p,t_0) \simeq \a^2  \left(H_1\over M_P\right)^2 
\left(\overline m \over H_{\rm eq}\right)^{1\over 2}  \left(p\over
p_1\right)^2, ~p<p_M.
 \label{13}
\eeq
As anticipated, it only depends on the three parameters $\a,m$ and
$H_1$ ($H_{\rm eq}\sim 10^{-55}M_P$ is the known value of the
curvature scale at the equilibrium epoch). 

In order to seed the galactic dynamo \cite{2} we must now require that
the above electromagnetic spectrum extends in frequency to include at
least the intergalactic scale $P_G(t_0)= (1 {\rm Mpc})^{-1}\simeq
10^{-14}$ Hz, namely
\beq
p_M=p_1(\overline m/H_1) \gaq p_G. 
\label{14}
\eeq
In addition, the fraction $r$ of seed energy density stored in the mode
$p_G$, relative to the CMBR energy density, $\Om_{\rm CMB}$, at the
epoch of galaxy formation, when $a_{\rm gal} \sim 10^{-2}a_0$, must
satisfy the (conservative) condition \cite{2}
\beq
r=\Om_A(p_G,t_{\rm gal})/\Om_{\rm CMB}(t_{\rm gal}) \gaq 10^{-34},
\label{15}
\eeq
in order that the produced magnetic fields may be large enough to seed
the galactic dynamo. Finally, the graviphoton mass cannot be
arbitrarily large, as the perturbation spectrum, integrated over all
modes, has to be at least smaller than one to avoid overcritical density,
and to avoid a Universe overdominated by the fluctuations of a massive
non-relativistic vector. This requires
\beq
\int_0^{p_M} d \ln \Om_\phi(p,t_0) \laq 1,
\label{16}
\eeq
where $\Om_\phi = \Om_A/\a^2$. 

An explicit computation shows that the lower bound on $m$ 
from eq. (\ref{14}) is always weaker than from eq. (\ref{15}), provided
we limit to ``realistic" values of the inflation scale, $H_1 \laq
10^{-1}M_P$. Assuming that this is indeed the case, we are left with
the conditions (\ref{15}) and (\ref{16}) which, by referring the  mass
to the physically interesting TeV scale, can be written explicitly as
\beq
10^{-11}{(1-\a^2)^{1\over 2}\over \a^4} \left(M_P\over H_1\right)^2
\laq 
 \left(m\over 1 {\rm TeV}\right)\laq 10^5 (1-\a^2)^{1\over 2}.
\label{17}
\eeq
The above bounds are illustrated in Fig. 1 for three different values of
the inflation scale. The allowed region lies below the curve of critical
density, and above the lines of seed magnetic energy. The allowed
windows close completely for $\log \a \laq -3.5$ and, in the limit $\a \ra
1$, for $H_1 \laq 10^{-8}M_P$. The standard inflation scale $H_1 \sim
10^{-5}M_P$ is not excluded, but the allowed region is larger for higher
scales. 

We may note, to this purpose, that de Sitter inflation at a nearly
Planckian scale is not necessarily ruled out by graviton production,
provided it is preceeded by a phase of superinflationary expansion
\cite{18}. In addition, the presence of such a superinflationary phase
leaves unchanged the constraints obtained with our previous analysis,  
if the de Sitter epoch is long enough to amplify the galactic scale $p_G$,
which is the relevant scale in the context of seed production.  This may
occur even for $H_1 \sim 0.1 M_P$  without clashing with obervational
data\cite{19,20}, provided -- as can be easily checked using the results
of \cite{18} --  $p_G (0.1 M_P/\overline m) \gaq 10^{-7}$Hz, i.e. 
provided $(m/1 {\rm TeV})\laq 10^8 (1-\a^2)^{1/2}$, which is always
automatically satisfied because of (\ref{17}). 

The vector masses allowed by the cosmological bounds (\ref{17}) thus
range from $10^5$ TeV for $\a \ll1$, to the KeV scale for $\a \sim1$ and
$H_1\ra 0.1M_P$ (smaller masses are not forbidden by eq. (\ref{17}),
but would require a fine-tuning of $\a$ to $1$). It becomes crucial, at
this point, to include in the discussion all existing experimental bounds
on the mixing parameter $\a$, for such a range of masses. 

\begin{figure}[t]
\begin{center}
\mbox{\epsfig{file=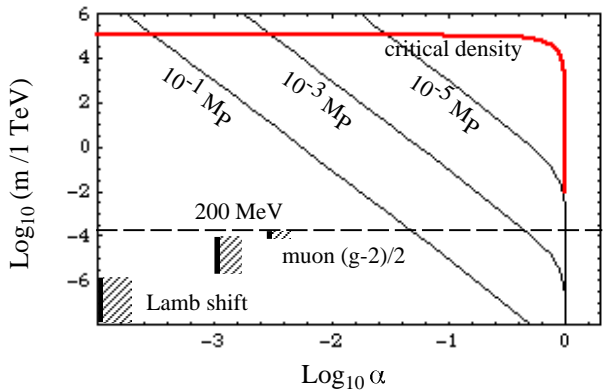,width=82mm}}
\vskip 5mm
\caption{\sl Allowed region for the inflationary production of magnetic
seeds, in the parameter space of the photon-graviphoton mixing. The
allowed region is below the bold curve (critical density), above
and to the right of the thin lines (seed magnetic energy, plotted for
different $H_1$), and above the $200$ Mev dashed line (determined by
the anomalous contribution to the $(g-2)/2$ ratio of the muon, by the
Lamb shift, and by other particle physics effects not shown in the
picture).}   \end{center}
\end{figure}

Known bounds, at present, are mainly referred to a graviphoton
interpretation of $V_\mu$  as the ``fifth-force" vector, coupled to
baryon number with gravitational strength. In that case, the mixing is
constrained on a macroscopic scale by geo-electric and geo-magnetic
data \cite{12}, leading to the condition $(m/1 {\rm TeV})\gaq 10^{-18}
\a (1-\a^2)^{-1/2}$, which is weaker, however, than the lower bound of
eq. (\ref{17}) (except in the limit $\a \ra 1$, where it forbids arbitrarily
small values of the vector mass). A much stronger bound on $\a$ would
follow \cite{12} from Cavendish's tests of Coulomb's law \cite{21}, but
it only applies when the graviphoton interaction is active on a
macroscopic range ($\la \gaq 1$ cm, i.e. $m \laq 10^{-4}$eV), and
then outside the mass range of Fig. 1. 

From $10^{-4}$ eV to $280$ MeV there are, however, important bounds
on the mixing obtained from nuclear physics \cite{13}. In particular: $\a
\laq 10^{-6}$ for $m=1.8$ MeV (from beam-dump experiments); $\a
\laq 10^{-4}$ for $m\leq 1$ MeV (from the Lamb shift in hydrogen
atoms); $\a
\laq 10^{-3}$ up to $m=100$ MeV, and $\a
\laq 10^{-2.5}$ for $m=200$ MeV (from the anomalous magnetic
moment of the muon). According to these bounds, all masses smaller
than $200$ Mev are associated to a mixing parameter $\a$ too small to
be compatible with the constraints (\ref{17}), and are thus to be
excluded from the allowed region of parameter space (see Fig.1, where
we have only reported the most significant bounds). 

Summing up the above results, we may conclude that the mixing of the
photon with a massive vector, gravitationally coupled to matter, can
provide, in principle, an efficient mechanism for the inflationary
generation of magnetic seeds for the galactic dynamo. A first estimate
of the allowed region in parameter space shows that such a mechanism
may be effective for a quite large mass window, $200$ MeV -- $10^5$
Tev. Higher inflation scales (typical, for instance, of string cosmology
\cite{6}) seem to be favoured, but also the standard inflation scale
$H_1 \sim 10^{-5}M_P$ is not excluded. The allowed mass window, in
particular, is well compatible with all conservative theoretical
estimates. The required mixing parameter, on the contrary, is
constrained by the severe bound $\log \a \gaq -3.5$, which is hardly
compatible with the most natural expectations based on extended
supersymmetric models, but not excluded in the context of
higher-dimensional unification schemes.

\acknowledgments
It is a pleasure to thank Massimo Giovannini and Gabriele Veneziano for
helpful discussions.

\end{document}